\documentclass{aa}

\usepackage{graphicx}
\usepackage{txfonts}
\begin{document}

   \title{Extended ammonia observations towards the `Integral-Shaped Filament'}


   \author{Gang Wu
          \inst{1,2,3,4}
          \and
          Keping Qiu\inst{2,5}
          \and
          Jarken Esimbek\inst{1,3}
          \and
          Xingwu Zheng\inst{2}
          \and
          Christian Henkel\inst{6,7}
          \and
          Dalei Li\inst{1}
          \and
          XiaoHong Han\inst{1}
          }

   \institute{Xinjiang Astronomical Observatory, Chinese Acadmy of Sciences,
              Science 1-street 150, Beijing Road, Urumuqi, China
              \email{wug@xao.ac.cn}
         \and
             School of Astronomy and Space Science, Nanjing University, 163 Xianlin Avenue, Nanjing 210023, China
         \and
             Key Laboratory of Radio Astronomy, Chinese Academy of Sciences, Urumqi 830011, P. R. China
         \and
             University of the Chinese Academy of Sciences, Beijing 100080, P. R. China
         \and
             Key Laboratory of Modern Astronomy and Astrophysics (Nanjing University), Ministry of Education, Nanjing 210023, China
         \and
             Max-Planck-Institut f\"{u}r Radioastronomie, Auf dem H\"{u}gel 69, 53121, Bonn, Germany
          \and
             Astronomy Department, King Abdulaziz University, PO Box 80203, Jeddah 21589, Saudi Arabia
             }

   \date{Received September 15, 1996; accepted March 16, 1997}


  \abstract
  {Recent observations suggest a scenario in which filamentary structures in the interstellar medium represent the first step towards clumps/cores and eventually star formation. The densest filaments would then fragment into prestellar cores owing to gravitational instability.}
  { We seek to understand the roles filamentary structures play in high-mass star formation.}
  {We mapped the integral-shaped filament (\object{ISF}) located at the northern end of the Orion A molecular cloud in NH$_{3}$ (1, 1) and $(2, 2)$.  The observations were made using the 25 meter radio telescope operated by the Xinjiang Astronomical Observatory, Chinese Academy of Sciences. The whole filamentary structure, about $1.2^{\circ } \times 0.6 ^{\circ }$, is uniformly  and fully sampled. We investigate the morphology, fragmentation, kinematics, and temperature properties in this region.}
  {We find that the morphology revealed by the map of velocity-integrated intensity of the NH$_{3}$ (1, 1) line is closely associated with the dust ridge revealed by the Herschel Space Observatory. We identify 6 "clumps" related to the well known OMC-1 to 5 and 11  "sub-clumps" within the map. The clumps and sub-clumps are separated not randomly but in roughly equal intervals along the \object{ISF}. The average spacing of clumps is 11.30'$\pm$1.31' (1.36$\pm$0.16 pc ) and the average spacing of sub-clumps is  7.18'$\pm$1.19' (0.86$\pm$0.14 pc).
  These spacings agree well with the predicted values of the thermal (0.86 pc) and turbulent sausage instability (1.43 pc) by adopting a cylindric geometry of the \object{ISF} with an inclination of $60^{\circ}$ with respect to the line of sight. We also find a velocity gradient of about 0.6 km s$^{-1}$ pc$^{-1}$  that runs along the \object{ISF} which likely arises from an overall rotation of the Orion A molecular cloud. The inferred ratio between rotational and gravitational energy is well below unity. Furthermore, fluctuations are seen in the centroid velocity diagram along the \object{ISF}. The OMC-1 to 5 clouds are located close to the local extrema of the fluctuations, which suggests that there exist gas flows associated with these clumps in the \object{ISF}.  The derived NH$_{3}$ (1, 1) and (2, 2)  rotation temperatures in the OMC-1 are about 30-40\,K while lower temperatures (below 20\,K) are obtained in the northern and southern parts of the \object{ISF}. In OMC-2, OMC-3, and the northern part of OMC-4, we find higher and lower temperatures at the boundaries and  in the interior, respectively.}
  {}


  \maketitle
%

\section{Introduction}
\label{introduction}

   In spite of the crucial role which high-mass stars (>8 M$_{\odot}$) play in the evolution of the Universe, their formation and early evolution remains poorly understood \citep{2007ARA&A..45..481Z, 2014prpl.conf..149T}. High-mass stars always form in stellar clusters embedded in Giant Molecular Clouds (GMCs). Recent observational studies reveal filamentary structures to be omnipresent in molecular clouds \citep[e.g.,][]{2010A&A...518L.102A, 2010ApJ...719L.185J, 2014A&A...567A..10L}. Moreover, according to a  "globular-filament" scenario of star formation \citep[][]{1979ApJS...41...87S}, filamentary structures in the interstellar medium represent the first step towards precluster clumps/prestellar cores and eventually star formation \citep[][]{2014prpl.conf...27A}. The densest filaments would then fragment into clumps/cores owing to gravitational instability \citep[e.g.,][]{2010ApJ...719L.185J}.

   Theories and models for filaments or gas cylinders have been studied for more than six decades \citep[e.g.,][]{1953ApJ...118..116C, 1964ApJ...140.1056O, 1992ApJ...388..392I}.
   The instability of gas cylinders shows some differences with respect to three-dimensional (3D) Jeans collapse. In a spherical gravitational Jeans collapse, perturbations of all wavenumbers grow at the same rate. However, in gas cylinders, perturbations of certain wavenumbers grow faster than others \citep[][]{1953ApJ...118..116C, 2010ApJ...719L.185J}.  Therefore, overdensities will tend to form at a length scale corresponding to this fastest growing wavenumber. Observationally, overdensities along filaments should be found at roughly regular intervals (the "sausage"  instability) \citep[e.g.,][]{2010ApJ...719L.185J}. Filaments are promising sites to study and test the physics of Molecular Cloud (MC) formation and fragmentation and to evaluate the earliest stages of high-mass star formation.

   The Orion complex is the nearest and probably best studied on-going star formation region that continues to produce both low- and high-mass stars \citep[][]{2008hsf1.book..459B}. At a distance of about 414$\pm$7 pc \citep[][]{2007A&A...474..515M}, it can be observed with good linear resolution even with a radio telescope of modest size. Meanwhile it is located about 15 degrees below the Galactic plane, leading to a less confused background than that typically encountered along the Galactic plane. Orion A is the largest MC ($\sim$31.5 deg$^{2}$) in the Orion complex \citep[][]{2005A&A...430..523W}. The `integral-shaped filament' (\object{ISF}) is the compact ridge at the northern end of the cometary Orion A cloud \citep[][]{1987ApJ...312L..45B, 1999ApJ...510L..49J}. The \object{ISF} contains several clumps, namely OMC-1, OMC-2, OMC-3, OMC-4, and OMC-5 \citep[e.g.,][]{1999ApJ...510L..49J,2006ApJ...653..383J}. The \object{ISF} is an active high-mass star formation site and has spawned a sequence of stellar groups of different ages in the past few million years and likely contains hundreds of young stellar objects \citep[][]{2008hsf1.book..459B}.

   To obtain physical information from the MCs, ammonia (NH$_{3}$) inversion transitions are an invaluable spectroscopic tool due to the molecule's  lack of depletion, its hyperfine structure, and its sensitivity to kinetic temperature \citep[][]{1983ARA&A..21..239H}. NH$_{3}$ (1, 1) and (2, 2) have been proved to be an excellent thermometer under ~20 K. It can also be a good thermometer for higher temperatures after some modification but with a reduction in precision \citep[][]{1983A&A...122..164W, 1988MNRAS.235..229D, 2004A&A...416..191T}. Moreover, the critical densities of NH$_{3}$ (1, 1) and (2, 2) are about 10$^{3}$ cm$^{-3}$ \citep[][]{1999ARA&A..37..311E, 2015PASP..127..299S}, thus providing a proper  tracer for the dense part of the filaments and their fragmentation.

   In this paper, we therefore present uniform and fully sampled maps of the \object{ISF} in NH$_{3}$ (1, 1) and (2, 2) and discuss the morphology, fragmentation, kinematics, and temperature properties in this region.


\section{Observations and database archives}
\subsection{NH$_{3}$ observations}
\label{observations}
The 25 meter radio telescope, operated by the Xinjiang Astronomical Observatory, Chinese Academy of Sciences, was used for all the NH$_{3}$ (1, 1) and (2,2) observations presented here. The rest frequency was set at 23.708564\,GHz for observing NH$_{3}$ (1, 1) at 23.694495\,GHz and NH$_{3}$  (2, 2) at 23.722633\,GHz, simultaneously. At this frequency, the telescope has a primary beam width (FWHM) of 125" obtained from point-like continuum calibrators and a velocity resolution of  0.098 km s$^{-1}$ provided by an 8192 channel Digital Filter Bank in a 64\,MHz bandwidth mode. All line intensities reported here are in units of main beam brightness temperature $T_{\rm mb}$.  $T_{\rm A}^{*}$ values were calibrated against periodically (6 s) injected signals from a noise diode. Due to a hardware problem, we additionally used GBT (Green Bank Telescope) NH$_{3}$ (1, 1) data \citep[][]{2017ApJ...843...63F} to calibrate the spectra to the $T_{\rm mb}$ scale with $\sim$ 14\% accuracy; see Appendix \ref{appCal}.

The telescope pointing and tracking accuracy is better than 18". A 22 - 24.2\,GHz dual polarization channel superheterodyne receiver was used as frontend. On a $T_{\rm A}^{*}$ scale, the typical system temperature is about 50 K at 23\,GHz. The map was made using the on-the-fly (OTF) mode with 6'x6' grid size and 30" sample step. All the observations were carried out in January 2014 under excellent weather conditions and above an elevation of 20$^{\circ}$.

\subsection{Archival data}
\label{archival}

The Herschel Space Observatory's 100 $\mu$m, 250 $\mu$m, and 500 $\mu$m images using SPIRE/PACS parallel scan mode data were taken by \citet[][]{2015A&A...577L...6S}.
The PACS data and SPIRE data used here are level 2.5 products produced by the Herschel interactive processing environment (HIPE) software, version 12.1 \citep[][]{2010ASPC..434..139O}. The Montage\footnote{This research made use of Montage. It is funded by the National Science Foundation under Grant Number ACI-1440620, and was previously funded by the National Aeronautics and Space Administration's Earth Science Technology Office, Computation Technologies Project, under Cooperative Agreement Number NCC5-626 between NASA and the California Institute of Technology.} and APLpy toolkits were used for cutout and plotting.

The archival $^{13}$CO spectra were observed with the IRAM 30 m telescope at Pico Veleta, Spain, in March, April, and October 2008. The data were obtained with
the HEterodyne Receiver Array (HERA, 9 dual polarisation pixels arranged in the form of a 3$\times$3 array with 24" spacing) with an individual beam width of about 11". The main beam efficiency at 220.4\,GHz was 0.545. The VErsatile Spectrometric and Polarimetric Array (VESPA) was used as backend providing 320\,kHz wide channels, which corresponds to roughly 0.4 km s$^{-1}$. The map was made in the on-the-fly (OTF) mode, with 5" data sampling in right ascension, and with steps of 12" in declination. The rms per channel is about 0.2 K on a main beam brightness temperature scale \citep[for more details, see ][]{2014ApJ...795...13B}.

\section{Data reduction and results}
\label{results}
The CLASS and GREG packages of GILDAS\footnote{http://www.iram.fr/IRAMFR/GILDAS/}, and also  python plot packages matplotlib \citep{2007CSE.....9...90H} and APLpy\footnote{http://aplpy.github.com} were used for all the data reduction. Second-order polynomial fitting is used for baseline removal and the mean rms per channel is about 0.3 K on a main beam brightness temperature scale. In order to convert hyperfine blended line widths to intrinsic line widths in the NH$_{3}$ inversion spectrum \citep[e.g.,][]{1998ApJ...504..207B}, we fitted the averaged spectrum using the GILDAS built-in `NH$_{3}$(1, 1)' fitting method which can fit all 18 hyperfine components simultaneously.

\begin{figure*}
   \centering
   \includegraphics[width=0.4\textwidth]{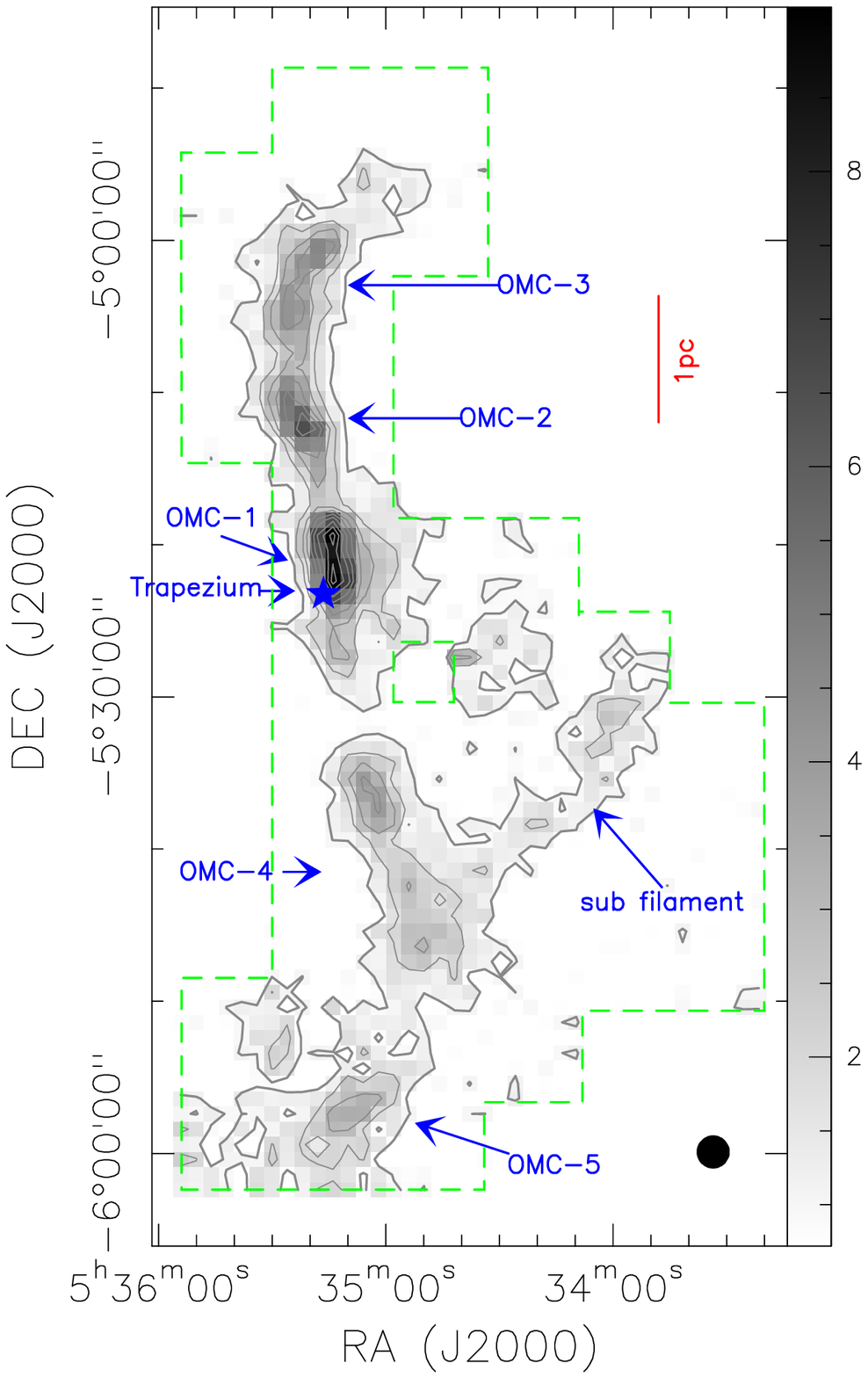}
   \includegraphics[width=0.4\textwidth]{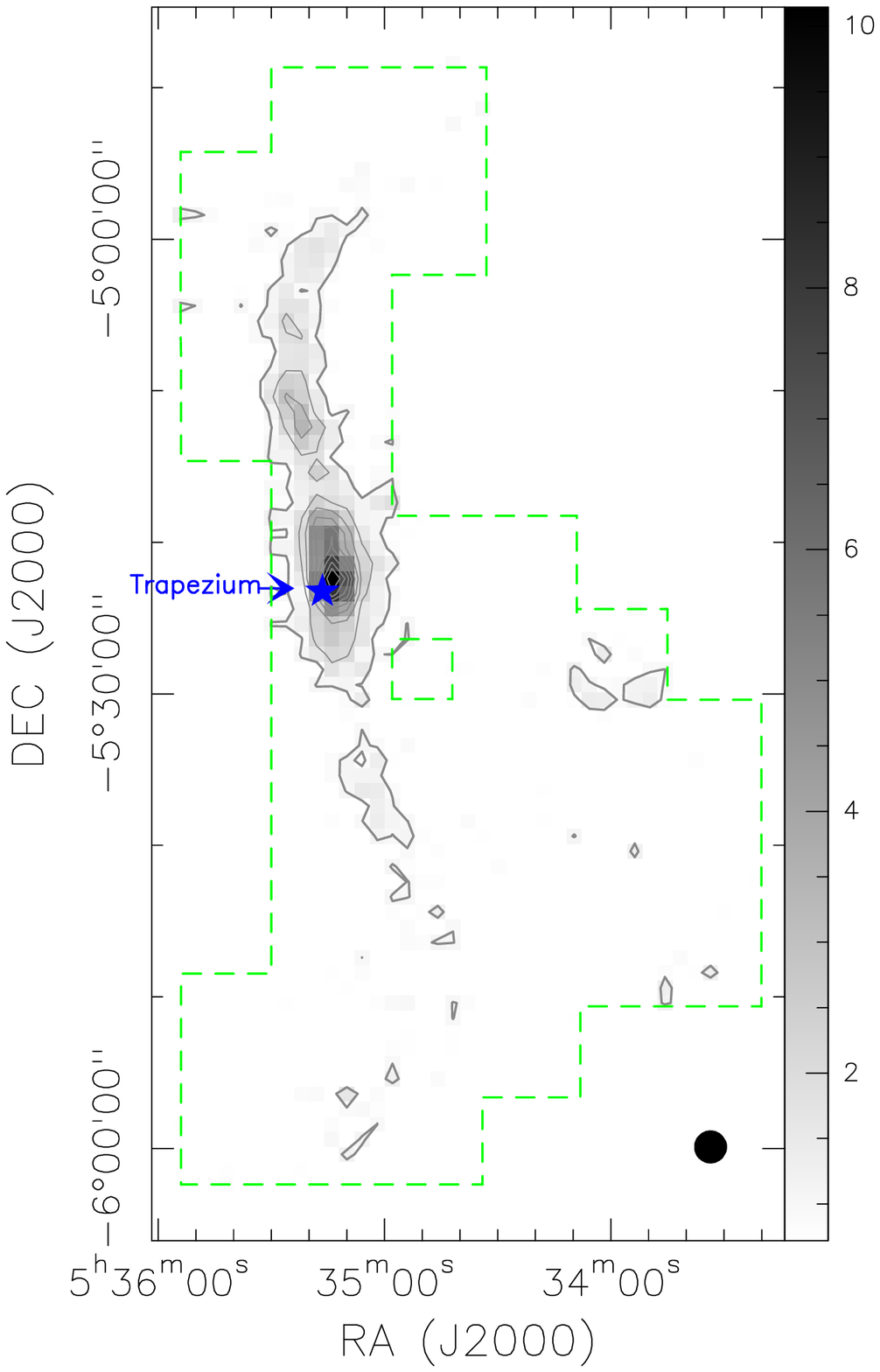}
   \caption{Integrated intensity (zeroth moment) maps of NH$_{3}$ (1, 1) (left) and (2, 2) (right). The integration range of each panel is 6.5 < Vlsr < 12.5 km s$^{-1}$. Contours start at 0.96 K km s$^{-1}$ (4$\sigma$) on a main beam brightness temperature scale and go up in steps of 0.96 K km s$^{-1}$. The limits of the mapped region are indicated with green dashed lines. A blue star in each panel indicates the Trapezium cluster and black filled circles in the lower right illustrate the half-power beam size.}
   \label{FigMom0}%
    \end{figure*}

The NH$_{3}$ (1, 1) and (2, 2) velocity-integrated line intensity (zeroth moment) maps are presented as gradations of gray-scale intensities and contours in Fig. \ref{FigMom0}. The integration range is from 6.5 to 12.5 km s$^{-1}$ to cover the main group of hyperfine components \citep[$\Delta$F = 0,][]{1983ARA&A..21..239H} of the NH$_{3}$ (1, 1) or (2, 2) transitions. We adopt this integration range since in relatively diffuse parts of the \object{ISF}, where NH$_{3}$  is barely detectable, the satellite lines \citep[$\Delta$F = $\pm1$,][]{1983ARA&A..21..239H} are usually not accessible and the velocity integrated intensity including satellite lines would lead to lower signal-to-noise (S/N) ratios. Meanwhile, we also constructed a velocity integrated intensity image including all hyperfine components (the main group and the four groups of satellite lines). The morphology, especially in the denser regions, is similar to the  image only including the main line. Furthermore, the results related to clump and sub-clump identification (Sect. \ref{morphology}) are not changed.

In Fig. \ref{FigMom0}, contours start at 0.96 K km s$^{-1}$ (4 $\sigma$, thick gray line) and go up in steps of 0.96 K km s$^{-1}$ (thin gray lines). $\sigma$ is equal to $rms\times  \Delta V \times \sqrt{N\_channels }$, where $rms$ is the averaged baseline channel rms of all spectra exported from the baseline fitting with CLASS (being part of the GILDAS package), $\Delta V$ is the channel spacing, and $N\_channels$ is the channel number in the integrated velocity range (6.5 to 12.5 km s$^{-1}$). In each panel, the Trapezium cluster is labeled as a blue star and the half-power beam width is illustrated as a black filled circle in the lower right bottom. The limits of the mapped region are indicated with green dashed lines. The clumps OMC-1 to 5 are labeled in the left panel. The red line in the top right of the left panel illustrates the 1 pc scale at a distance of 414 pc \citep[][]{2007A&A...474..515M}. An additional "sub filament" structure in the west of OMC-4 is also labeled in the left panel. The whole map consists of a total of 45240 raw spectra. After smoothing to 1' $\times$ 1' areas, 1764 uniformly spaced spectra remain, covering a region about 1.2 degrees in declination and about 0.6 degrees in right ascension.

In the left panel of Fig. \ref{FigMom0}, the zeroth moment map of NH$_{3}$ (1, 1) shows an integral-shaped  morphology ($\int$) as observed in $^{13}$CO \citep[see][]{1987ApJ...312L..45B}. Compared to $^{13}$CO, NH$_{3}$ (1, 1) highlights the main filamentary structure and the fragmentation  in a clearer way and we find, unlike what is seen in $^{13}$CO, little diffuse gas around the \object{ISF}. In NH$_{3}$ (1, 1), the length of the \object{ISF} is about 7 pc, and the width is about 1 pc at a distance of 414 pc. The densest regions all show elongated morphologies, mainly along the main axis of the \object{ISF}, hinting at the existence of sub-structures which are displayed in more detail in Fig. 7 of \citet[][]{2017ApJ...843...63F}. A separated emission region located between the OMC-1 and the already mentioned western sub-filament is not analysed in this work, since only a few pixels show NH$_{3}$ (1, 1) detections in this region. According to the right panel of Fig. \ref{FigMom0}, most of the NH$_{3}$ (2, 2) line detections are obtained in the northern part of the \object{ISF}. Considering weaker NH$_{3}$ (1, 1) intensities in the southern part of the \object{ISF} (left panel of Fig. \ref{FigMom0}) and keeping in mind that CO is also weaker in the south \citep[e.g.,][]{1987ApJ...312L..45B, 2005A&A...430..523W}, this is likely caused by lower NH$_{3}$  column densities in the southern part of the \object{ISF}.

\section{Discussion}
\subsection{Ammonia morphology and two level fragmentation in the \object{ISF}}
\label{morphology}

A false color Herschel infrared image of the \object{ISF} is shown in the left panel of Fig. \ref{FigHerschel} (red for 500 $\mu$m, green for 250 $\mu$m, and blue for 100 $\mu$m), overlaid with NH$_{3}$ (1, 1) integrated intensity contours as in Fig. \ref{FigMom0}. From this panel, we can see the contours are closely associated with the dust ridge presented by the Herschel infrared emission. As mentioned in Sect. \ref{results}, there is a sub-filament in addition to the main filamentary structure, which is extruded in the west of OMC-4 and which is identified for the first time in NH$_{3}$.
The sub-filament has a length of about 20', that is, about 2 pc at a distance of 414 pc. In the left panel of Fig. \ref{FigHerschel}, there is also an obvious infrared feature corresponding to this sub-filament.

From the NH$_{3}$ (1, 1) integrated intensity map, we find that the six regions with strongest NH$_{3}$ emission are located along the \object{ISF}.  In the right panel of Fig. \ref{FigHerschel}, we marked the six  regions (two related to OMC-4) with strongest NH$_{3}$ emission as "clumps"  with red crosses on the NH$_{3}$  map.
The emission peaks correspond to significant star-forming regions within the Orion A Molecular Cloud \citep[e.g., ][]{1987ApJ...312L..45B, 1999ApJ...510L..49J, 2006ApJ...653..383J}.
Following the terminology of \citet[][]{1999ApJ...510L..49J, 2006ApJ...653..383J}, we name them OMC-1 to 5. Since the two regions in OMC-4 are separated by more than a typical clump size of 1 pc \citep[e.g., hierarchical fragmentation in][]{2009ApJ...696..268Z, 2014MNRAS.439.3275W}, we consider them as two separate "clumps". These clumps are not randomly spaced but at roughly equal intervals. The average spacing of these five clumps is 11.30'$\pm$1.31' (given uncertainties are errors of the mean), that is, 1.36$\pm$0.16 pc  at a distance of 414 pc \citep[][]{2007A&A...474..515M}.
We then used the Clumpfind2d algorithm \citep[][]{1994ApJ...428..693W} with a threshold of 5$\sigma$ and increments of 2$\sigma$ ($\sigma$ is defined in Sect. \ref{observations}) to identify sub-structure. \citet[][]{2014ApJ...790...84L} and \citet[][]{, 2016ApJ...833..209O}, for example, used similar parameters to identify sub-structures. Eleven such sub-structures were identified within the NH$_{3}$  map after excluding a fake one located at the boundary of the map. We also present the pixels used to define these sub-structures in Appendix \ref{appSub}. As the sub-structures are smaller than 1 pc and larger than 0.1 pc, we name these sub-structures "sub-clumps".
\citet[][]{2009ApJ...699L.134P} argued the specific selection of the parameters in the Clumpfind2d algorithm may result in slight differences in identifying the condensations, but it works well for relatively isolated sub-clumps which we find in the \object{ISF}. In our observation, the clumpfind2d algorithm is more sensitive to the threshold we have chosen, and not to the increments. Using a threshold of 3$\sigma$, the procedure tends to find some additional artificial condensations at the edge of the image. Therefore, we adopt a robust threshold of 5$\sigma$ and increments of 2$\sigma$  leading to an identification of those sub-clumps that one would tend to select by eye.  We marked these sub-clumps as yellow filled circles in the right panel of Fig. \ref{FigHerschel}. As we can see, the ten sub-clumps  within the main structure of the \object{ISF} are also regularly separated. The average spacing is 7.18'$\pm$1.19', which corresponds to 0.86$\pm$0.14 pc at 414 pc.

The results may suffer from the  effect of the small number statistics. Before we interpret the results, we considered 1 million tests of six-element random arrays in the range of 0 to 1.36$\times$6 pc. About 1.1\% of these random arrays are more equally spaced (the ratio of the standard deviation of the mean to the mean value is less than 0.16 pc/1.36 pc). Subsequently, we considered 1 million tests of ten-element random arrays in the range of 0 to  0.86$\times$10 pc. About 2.5\% of these random arrays are more equally spaced (the ratio of the standard deviation of the mean to the mean value is less than 0.14 pc/0.86 pc).

The approximately equal intervals of the clumps can be explained by a so-called sausage instability firstly proposed by \citet[][]{1953ApJ...118..116C} for an incompressible gas cylinder. Following \citet[][]{2010ApJ...719L.185J}, for an isothermal cylinder with infinite radius, the characteristic length between adjacent condensations is about $22H$. $H$ is the isothermal scale height given by $H= c_{s}  (4  \pi G  \rho_{c} )^{-1/2}$, where c$_{s}$ is the sound speed, G is the gravitational constant, and $\rho_{c}$ is the gas mass density at the center of the filament. For an isothermal cylinder of finite radius $R$ surrounded by an external, uniform medium, the characteristic length for R $\gg H$ is also about $22H$. If $R \ll H$, the characteristic length becomes about $11R$.

We assumed the \object{ISF} is an isothermal cylinder with a typical kinetic temperature of 18 K (see Sect. \ref{temperature}) and a central volume average density of the \object{ISF} of $10^{4}$ cm$^{-3}$ \citep[][]{1987ApJ...312L..45B, 1994A&A...281..209C, 1999ApJ...510L..49J}. Then the sound speed c$_{s}$ = $\sqrt{kT/\mu_{H}m_{H}} = 0.25$ km s$^{-1}$, taking $\mu_{H} = 2.4$ and $m_{H}$ = 1.674$\times$10$^{-24}$ g. The isothermal scale height is  $H \sim$ 0.04 pc. Therefore, under the regime $R \gg H$, the characteristic length ($22H$) of the \object{ISF} is about 1 pc.  Assuming a median inclination of a randomly oriented filament of 60$^{\circ}$ with respect to the line of sight \citep[][]{1989ARA&A..27...41G, 1993ApJ...404L..83H}, the projected spacing will be 1 pc $ \times$  sin(60$^{\circ})$ $\sim$ 0.86 pc, which is exactly the averaged spacing of the sub-clumps (0.86$\pm$0.14 pc). As we can see, the inclination of the \object{ISF} is relevant in defining the separation of the clumps. We are collecting distance measurements in the entire Orion A filament to fit the inclination of the \object{ISF} in a separate paper. According to the preliminary result of  the fitting of 17 YSO distances derived by the VLBA paradox measurement \citep[][]{2017ApJ...834..142K}, the derived inclination is about 60 $^{\circ}$ and the northern part is more nearby (Wu et al., in preparation).
As suggested by \citet[][]{2010ApJ...719L.185J}, the above calculations assume that thermal pressure dominates the overall pressure budget. Because the observed line widths in most molecular clouds typically exceed the thermal sound speed, turbulent pressure usually dominates over thermal pressure. To consider the turbulence's contribution, we have to replace the sound speed by the typically observed velocity dispersion  $\sigma$. Therefore, adopting a FWHM line width of 1.0 km s$^{-1}$ for NH$_{3}$ (1, 1) (see sect. \ref{kinematics}), the velocity dispersion $\sigma$ is related to the FWHM line width $\Delta V$ as $\sigma = \Delta V / \sqrt{8 \ln(2)}$  = 0.42 km s$^{-1}$ for a Gaussian line profile. Then the effective scale height, replacing the previously defined $H$-parameter, is about 0.07 pc. The characteristic length then becomes about 1.65 pc. Assuming again that the inclination of the \object{ISF} is $60^{\circ}$, the projected spacing will be 1.65 pc $\times$  sin(60$^{\circ})$ $\sim$ 1.43 pc, which is in good agreement with the averaged separation of the clumps we find (1.36$\pm$0.16 pc). In short, we observe a two-level fragmentation in the \object{ISF}, which can be  explained by thermal and turbulent sausage instability, respectively. In this context it is important to note that the Jeans length $\lambda_{J}$  $\sim$  0.27 pc ($T=18 K $,  $n(H_{2})=10^{4}$ cm$^{-3}$) cannot explain this fragmentation.

The previous spacing calculations are all based on the typical temperature and velocity, which represent the  peak values of their statistics. These values may characterize the typical properties of the gas which is not directly affected by star formation. The median and mean line widths are 1.3 km $\rm s^{-1}$ and 1.5 km $\rm s^{-1}$ , respectively (see Sect. \ref{kinematics}) while median and mean kinetic temperatures are 23 K and 26 K, respectively (see Sect. \ref{temperature}). Adopting the median temperature and line width, the corresponding spacings for thermal  and turbulent cylinder fragmentation are 1.1 pc  and 2.18 pc. Adopting the means of temperature and line width, the corresponding spacings for thermal and turbulent cylinder fragmentation are 1.17 pc and 2.5 pc.
Therefore, for the median case, inclinations of  47$\rm ^{\circ}$ and 32$\rm ^{\circ}$ should be assumed to explain the observed spacings. For the mean case, inclinations of  51$\rm ^{\circ}$ and 38$\rm ^{\circ}$ are needed. Neither for the median nor the mean case, can the calculated thermal and turbulent spacings be explained by a consistent inclination.

The (quasi-) periodically spaced fragmentation in the \object{ISF} has been noticed by many previous studies from several tens of parsecs to a few percent  of a parsec
\citep[e.g.,][]{1991A&A...247L...9D, 1999ApJ...510L..49J, 2013ApJ...763...57T}. \citet[][]{1991A&A...247L...9D}  first  presented an obvious periodical density structure with C$^{18}$O and suggested the periodicity is a result of externally triggered and magnetically mediated cloud fragmentation. More recently, \citet[][]{2013ApJ...763...57T} summarized the fragmentation in the entire Orion complex from large to small scale, including a spacing of  $\sim$4.7$^{\circ}$ (34 pc) corresponding to the separation between the Orion A and B giant molecular clouds \citep[e.g.,][]{1986ApJ...303..375M}, a spacing of 9' - 10' (1.0 - 1.2 pc) corresponding to the separation of the large scale clumps, for example, OMC-1, OMC-2 \citep[e.g.,][]{1991A&A...247L...9D, 1993ApJ...404L..83H}, a  spacing of about 2.5' (0.3 pc) corresponding to the separation between small scale clumps \citep[e.g.,][]{1999ApJ...510L..49J}, and the spacing corresponding to the dense cores, for example, 17" (0.034 pc) in OMC-2/3 and 30" (0.06 pc) in OMC-1 \citep[e.g.,][]{1998ApJ...502..676W, 2013ApJ...763...57T, 2016A&A...587A..47T}. They argued that fragmentation spacings on scales of 0.1 - 1 pc are roughly consistent with thermal  fragmentation, while on small scales (< 0.1 pc) fragmentation spacings are below those of the thermal fragmentation. Our data showing two levels, that is,  1.36$\pm$0.16 pc and 0.86$\pm$0.14 pc, are consistent with turbulent and thermal cylinder fragmentation.

Because our NH$_{3}$ observations could well resolve the spacings between the clumps / sub-clumps, the spacings identified by our NH$_{3}$ observations  are reliable. Lower-resolution observations may more clearly present the larger-scale fragmentation \citep[i.e., Fig. 3 in][]{2016A&A...587A..47T}. Meanwhile, according to a high-fidelity map of G11.11-0.12, \citet[][]{2013A&A...557A.120K} suggested that on scales larger than 0.5 pc, fragmentation may still be affected by global instabilities which is in agreement with the model of a self-gravitating cylinder. Therefore, the two-level hierarchical cylinder fragmentation at scales of 0.86 and 1.36 pc   should be reasonable. Furthermore, this two-level fragmentation indicates that the clump separation is due to turbulent pressure confinement and the sub-clump separation is due to thermal pressure confinement; that is, turbulence should be dissipated from the scale of clumps to sub-clumps in the \object{ISF}.  We studied the line width at different scales in a quiescent clump, OMC-2, in Appendix \ref{appDis}. As the linear scale length decreases, the velocity dispersion decreases as is also deduced by many previous studies \citep[i.e.,][]{1998ApJ...504..223G,1998ApJ...504..207B}.

\subsection{Kinematics from NH$_{3}$}
\label{kinematics}

Figure \ref{FigVel} presents the kinematics of the \object{ISF} derived from NH$_{3}$ (1, 1). The left panel displays velocity (color image) and the right panel indicates the intrinsic line width (i.e., for an individual hyperfine component, gray image). They are all derived by the NH$_{3}$(1,1) fitting method in GILDAS. The two images are overlaid with  NH$_{3}$ (1, 1) integrated intensity contours as in Fig. \ref{FigMom0}. Firstly, we can see that a clear velocity gradient runs along the main part of the \object{ISF}. Following the procedure outlined in \citet[][]{1993ApJ...406..528G}, we fitted the velocity gradient as a simple linear form: $V_{LSR} = v_{0} + a\times\delta\alpha + b\times\delta\beta$, where $\delta\alpha$ and $\delta\beta$ are offsets in right ascension and declination (in radians), and $v_{0}$ presents the systemic velocity of the cloud. The velocity gradient is therefore given by  $\nabla v = (a^{2} + b^{2})^{0.5}/D $ at a distance of D.
The derived gradient is 0.6$\pm0.08$ km s$^{-1}$ pc$^{-1}$ at 414 pc using all the spectra with peak line flux larger than 5$\sigma$. This linear velocity fitting is based on a solid body approximation for the \object{ISF}. To check this assumption, we study the velocity residuals (V$_{\rm obs}$\,$-$\,V$_{\rm rigid}$) in Appendix \ref{appRes}. The velocity residuals do not present major departures from a linear velocity distribution.
The observed velocity gradient is  comparable to the overall rotation of the Orion A Molecular Cloud \citep[e.g.,][]{2005A&A...430..523W}.
\citet[][]{2005A&A...430..523W} suggested an increase in the inclination of the filament as a function of time with respect to the line of sight (see Sect. \ref{morphology}).
Taking the \object{ISF} as a homogeneous rigidly rotating cylinder, the ratio of rotational (around its center near $\delta \sim -5^{\circ }30'$) to gravitational energy is
\begin{equation}
\begin{aligned}
\beta = \frac{E_{rot}}{E_{grav}}=\frac{1}{2} \frac{ML^2}{12}(1+\frac{3}{2x^2}) \omega ^2 / \frac{3}{2}\frac{GM^2}{L}f(x)  \approx \frac{\omega ^2}{4\pi G\rho}\frac{L^2}{9R^2},
\end{aligned}
\end{equation}
where $G$ is the gravitational constant and $\rho$, $L$, $R$, and $\omega$ are the density, height, radius and the angular velocity of the cylinder. $x\equiv L/D$, $f(x)\sim 1$ \citep[][]{1992ApJ...400..579B}. \\
We can also write this as
\begin{equation}
\begin{aligned}
\beta = \frac{3.0 \times 10^{-3}  \omega_{-14}^{2}}{n_{4}}  \frac{L^2}{9R^2},
\end{aligned}
\end{equation}
where, $\omega_{-14}$ is the angular velocity in units of $10^{-14} s^{-1}$  and  $n_{4}$ is the gas density in units of $10^{4}$ cm$^{-3}$ \citep[][]{1984A&A...137..108M, 2013A&A...553A..58L}.  Adopting an angular velocity of $1.94 \times 10^{-14} s^{-1}$ and an average density of the \object{ISF} of $10^{4}$ cm$^{-3}$  \citep[][]{1987ApJ...312L..45B, 1991A&A...247L...9D, 2014ApJ...795...13B},  then $\beta$ $\sim$ 0.25. Thus, the rotational energy is a small fraction of the gravitational energy.

From the line width image, three patterns are apparent. Firstly, the most dispersed region is located around the Trapezium in OMC-1. Secondly, in  the northern part of the \object{ISF}, very low and also uniform intrinsic (for individual hyperfine components) dispersions ($\sigma \sim$ 0.4$\pm$0.08 km s$^{-1}$) are present. Very narrow line widths  of the NH$_{3}$ (1, 1) main line are mostly found in this region. Thirdly,  irregular and larger dispersion is present in the southern part of the \object{ISF}. This may mainly be because in this region S/Ns are not as high. We study the FWHM line widths of the NH$_{3}$ (1, 1) main lines with  a  peak line flux threshold of 5$\sigma$ as summarized in Fig. \ref{FigStaV}. The line width refers to individual hyperfine component and is fitted using the GILDAS built-in `NH$_{3}$(1, 1)' fitting method. Obviously the line width distribution of the NH$_{3}$ (1, 1) gas has an outstanding peak around 1 km s$^{-1}$. Therefore we adopted this line width  as the typical width for the fragmentation analysis in Sect. \ref{morphology}. The median and mean intrinsic line widths are 1.3 km s$^{-1}$ and 1.5 km s$^{-1}$, respectively.

To clearly show the kinematics along the \object{ISF}, we created the poly-line  position-velocity (PV) diagram presented in Fig. \ref{FigPV}. The loci of the PV diagram, which runs through the ridge of the  \object{ISF} and crosses the cores, are illustrated by the gray solid lines in the left panel of Fig. \ref{FigVel}. We averaged 5 pixels (5') perpendicular to the loci in order to present the integral kinematics and to make sure that the exact location of the poly-line, chosen by eye, is not a critical parameter. From the diagram, we can see that, (1) there is an overall velocity gradient along the \object{ISF} as shown in the velocity image (left panel of Fig. \ref{FigVel}); (2) bounded by OMC-1, the features of the northern part are coherent and compact, and the southern ones are diffuse and faint; and (3) the two sub-clumps in the OMC-1 show a large difference in velocity.

Filamentary structures of MCs may be seen as accretion channels to their associated clumps and lead to a seesaw velocity distribution \citep[e.g.,][]{2013ApJ...766..115K, 2015ApJ...804..141Z}. Therefore, to further examine the gas kinematics, Fig. \ref{FigVelDec} presents the centroid velocity obtained by the main lines of NH$_{3}$ (1,1). We firstly excluded the spectra with peak line flux less than 5$\sigma$ to avoid spurious results. We then averaged the spectra within a 1'-wide belt in declination along right ascension which is roughly perpendicular to the main filamentary structure. Finally, single Gaussian  fits were used to get the centroid velocities of these averaged spectra. The result is summarized in Fig. \ref{FigVelDec} with blue dots and lines. Beside the overall motion (velocity gradient), fluctuations are also seen along the \object{ISF}. The locations of OMC-1 to 5 are labeled in this figure and are all found near the local extrema of the fluctuations. This may hint at gas flows toward the clumps in the \object{ISF} but may also arise from  effects related to star formation. Besides, we also derive the centroid velocities of IRAM $^{13}$CO (2-1) data with the same method as for NH$_{3}$ and summarize this in Fig. \ref{FigVelDec} with red dots and lines. The large velocity shifts and fluctuations are roughly the same as for NH$_{3}$. However, there is a striking velocity jump ($\sim$ 2 km s$^{-1}$) of NH$_{3}$ relative to $^{13}$CO around OMC-1.
According to $^{13}$CO observations in Orion A \citep[e.g., ][]{1987ApJ...312L..45B, 2014ApJ...795...13B}, the northern part of the Orion A molecular cloud is a coherent structure. It is not likely that $^{13}$CO is from a separate group of clouds along the line of sight.
Considering the strong NH$_{3}$ emission originating from OMC-1's dense hot core with a radial velocity of $\sim5$ km s$^{-1}$, while that of the nearby Orion ridge, which dominates the $^{13}$CO emission, is near 8 km s$^{-1}$ \citep[e.g.,][]{1982ApJ...259L.103G, 1998ApJ...502..676W}, the discrepancy is readily explained.

\subsection{Temperature}
\label{temperature}

The lack of allowed radiative transitions between the K = 1 and 2 ladders of NH$_{3 }$ makes their relative populations highly sensitive to collisions, therefore making them a good estimator of the gas kinetic temperature. The rotation temperature of the two levels is derived following the method in \citet[][]{1983ARA&A..21..239H}.
The integrated optical depth of the main feature of the NH$_{3}$ (1, 1) transition, excluding the satellite lines, is derived by the GILDAS built-in `NH$_{3}$(1,1)' fitting method. Subsequently, we used the formula below to obtain the rotational temperature,
\begin{equation}
\begin{aligned}
& T_{rot}(2,2:1,1) = \\
&-41.5 \div \ln[\frac{-0.282}{\tau_{m}(1,1)} \ln{[1-\frac{\triangle T_{a}(2,2,m)}{\triangle T_{a}(1,1,m)}]}(1- \exp ^{-\tau_{m}(1,1)})],
\end{aligned}
\end{equation}

where $\triangle T$  is the main line intensity and $\tau_{\rm m}$ denotes the main group opacity derived using the GILDAS built-in `NH$_{3}$ (1, 1)' fitting.  $\triangle T$  of NH$_{3}$ (1, 1) and NH$_{3}$ (2, 2) are derived using the GILDAS built-in `GAUSS' fitting. The mean uncertainty of the derived temperature is 2K.

The color image in the left panel of Fig. \ref{FigTem} presents the rotational temperature ($T_{R}$) distribution of the \object{ISF} with overlaid NH$_{3}$(1, 1) integrated intensity contours as in Fig. \ref{FigMom0}. $T_{R}$ exceeds 40 K towards several positions. There the intensities of NH$_{3}$(2, 2) are even stronger than those of NH$_{3}$(1, 1) and they mainly come from OMC-1. As an example, nine spectra around the Trapezium are presented in the right-bottom panel of Fig. \ref{FigTem} (black spectra for NH$_{3}$(1, 1), red spectra for NH$_{3}$(2, 2)). From the left panel we can see, the gas around the Trapezium cluster is characterized by $T_{R} \sim$ 30-40 K. The gas in OMC-2, OMC-3, and the northern part of OMC-4 has a lower  $T_{R}$ of about 20K or even lower than that. In OMC-2 to 4, we can see higher  $T_{R}$ at the boundaries and lower  $T_{R}$ in the interior of these clumps. This may illustrate that heating  originates mainly from the outside and that these clumps may still be in an early evolutionary stage.

A statistical distribution of $T_{R}$ is summarized in the top-right panel of Fig. \ref{FigTem} and shows a typical value of about 16 K, corresponding to a kinetic temperature of 18 K \citep[following][]{2004A&A...416..191T}, which has been used in Sect. \ref{morphology} to calculate expected clump separations inside a cloud of cylindric morphology. The median and mean values are 19 K and 20.8 K, corresponding to a kinetic temperature of 23 K and 26 K \citep[following][]{2004A&A...416..191T}.

\subsection{A brief comparison with GBT data}
While this work was in progress, \citet[][]{2017ApJ...843...63F} published ammonia data of the \object{ISF} obtained with the Green Bank Telescope. This map,  taken with a resolution of 32", extends slightly further to the south but is limited to the narrow ridge and does not include the western protrusion of the filament we see between $\delta \sim -5^{\circ }25'$ and $\delta \sim -5^{\circ }40'$.  Agreement in the overall velocity, line width, and kinetic temperature distributions is fairly good. We also identify the cloud displaced 5'-10' eastwards of the backbone of the filament near $\delta \sim -5^{\circ }52'$.

\section{Summary}
\label{summary}

We have mapped the \object{ISF} in NH$_{3}$(1, 1) and (2, 2)  located at the northern end of the Orion A molecular cloud. The  uniformly and fully-sampled observations cover a region of about 1.2 degrees in declination and about 0.6 degrees in right ascension. The main conclusions of this work are as follows.

The NH$_{3}$ morphology of the \object{ISF} is associated with the Herschel infrared dust ridge and also shows an \object{ISF} morphology. A 2-pc-length `sub-filament' extruding toward the northwest of OMC-4 also shows a clear corresponding far infrared feature and is detected for the first time in NH$_{3}$.

We identify 6 clumps and 11 sub-clumps in the integrated intensity maps of NH$_{3}$(1,1). We find the clumps and sub-clumps are all separated not randomly but in roughly equal intervals. Comparing our data with (quasi-) periodically spaced fragmentation in the \object{ISF} revealed by previous studies, we propose that hierarchical fragmentation around the clump scale ($\sim$ 1 pc) shows two levels, that is, clumps at separations of 1.36$\pm$0.16 pc and sub-clumps with a spacing of 0.86$\pm$0.14 pc. Adopting an inclination of 60$^{\circ}$, this two-level fragmentation is consistent with turbulent and thermal fragmentation inside a gas cloud with cylindric morphology.

A velocity gradient of 0.6 km s$^{-1}$ pc$^{-1}$ runs along the \object{ISF}, likely due to an overall rotation of the Orion A cloud. Using this velocity, we compared rotational and gravitational energy of the \object{ISF}. The ratio is about 0.25, demonstrating that the rotational energy is a small fraction of the gravitational energy. We made an averaged centroid velocity diagram along the \object{ISF} to check the roles of filaments as the accretion channels to clumps.  From the diagram, fluctuations are seen  along the \object{ISF}. OMC-1 to 5 are all located near local extrema of the fluctuations. This may indicate gas flows toward the clumps in the \object{ISF} but  may  also be explained in terms of  star forming activity. Moreover, a striking velocity jump of NH$_{3}$  relative to $^{13}$CO around the OMC-1 is found in this diagram. The explanation is that NH$_{3}$ is enhanced by hot dense clumps while $^{13}$CO is mainly associated with less dense cooler gas.

The derived NH$_{3}$ (1,1) and (2,2) rotation temperatures in the OMC-1 are about 30-40 K. Lower rotation temperatures (below 20 K) are present at the northern and the southern parts of the \object{ISF}. OMC-2, OMC-3, and northern part of OMC-4 show larger temperatures at the boundaries and lower temperatures in the interior of these clumps. This may illustrate that heating mainly occurs from the outside and that these clumps may still reside in an early evolutionary stage.

\begin{acknowledgements}

    This work was funded by the National Natural Science foundation of China under grant 11433008, 11603063, and the Program of the Light in China's Western Region (LCRW) under grant Nos and 2015-XBQN-B-03, partly supported by the National Natural Science foundation of China under grant 11373062, 11303081. K.Q. acknowledges the support from National Natural Science Foundation of China (NSFC) through grants NSFC 11473011 and NSFC 11590781.

\end{acknowledgements}

   \begin{figure*}
   \centering
   \includegraphics[width=0.4\textwidth]{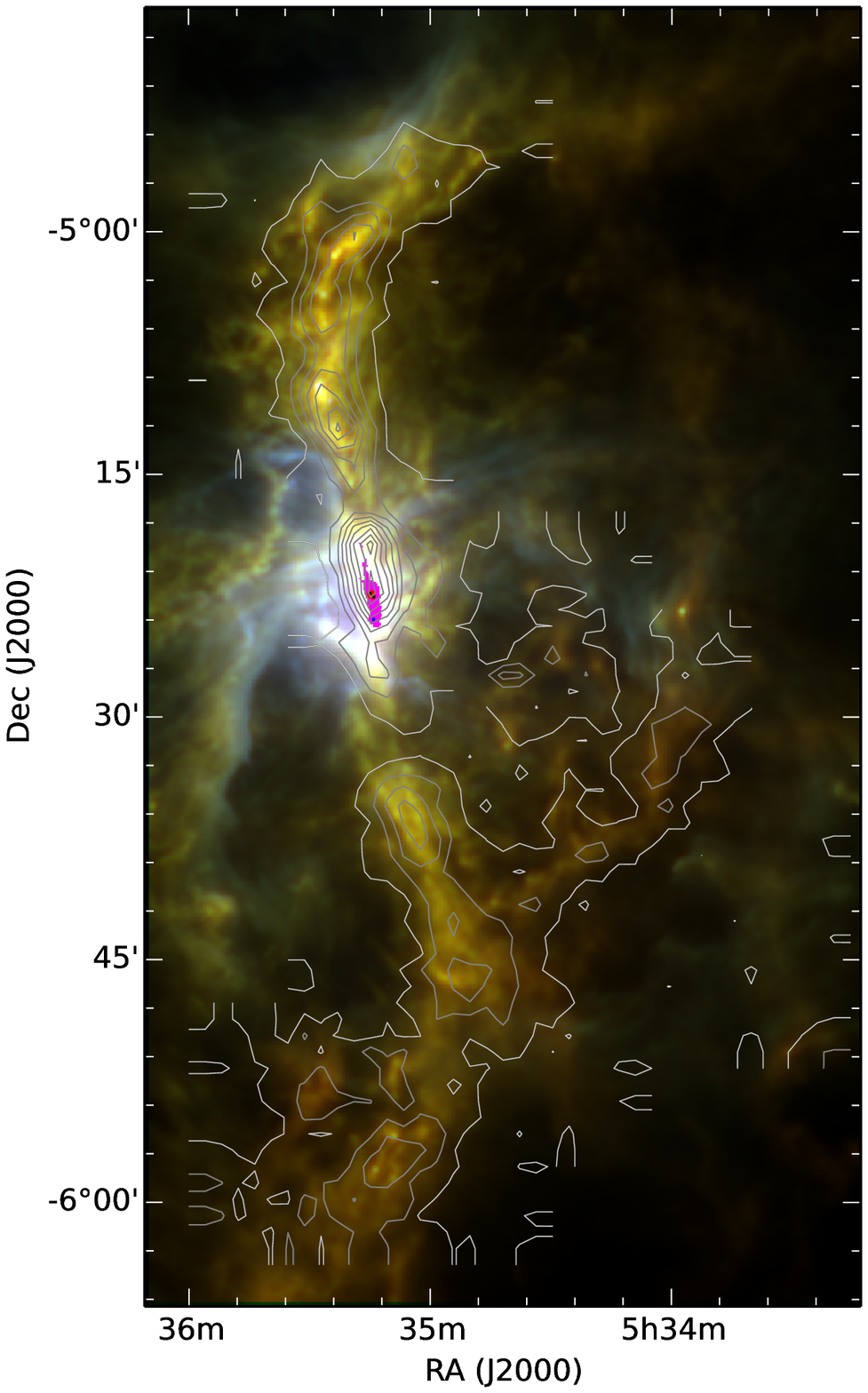}
   \includegraphics[width=0.4\textwidth]{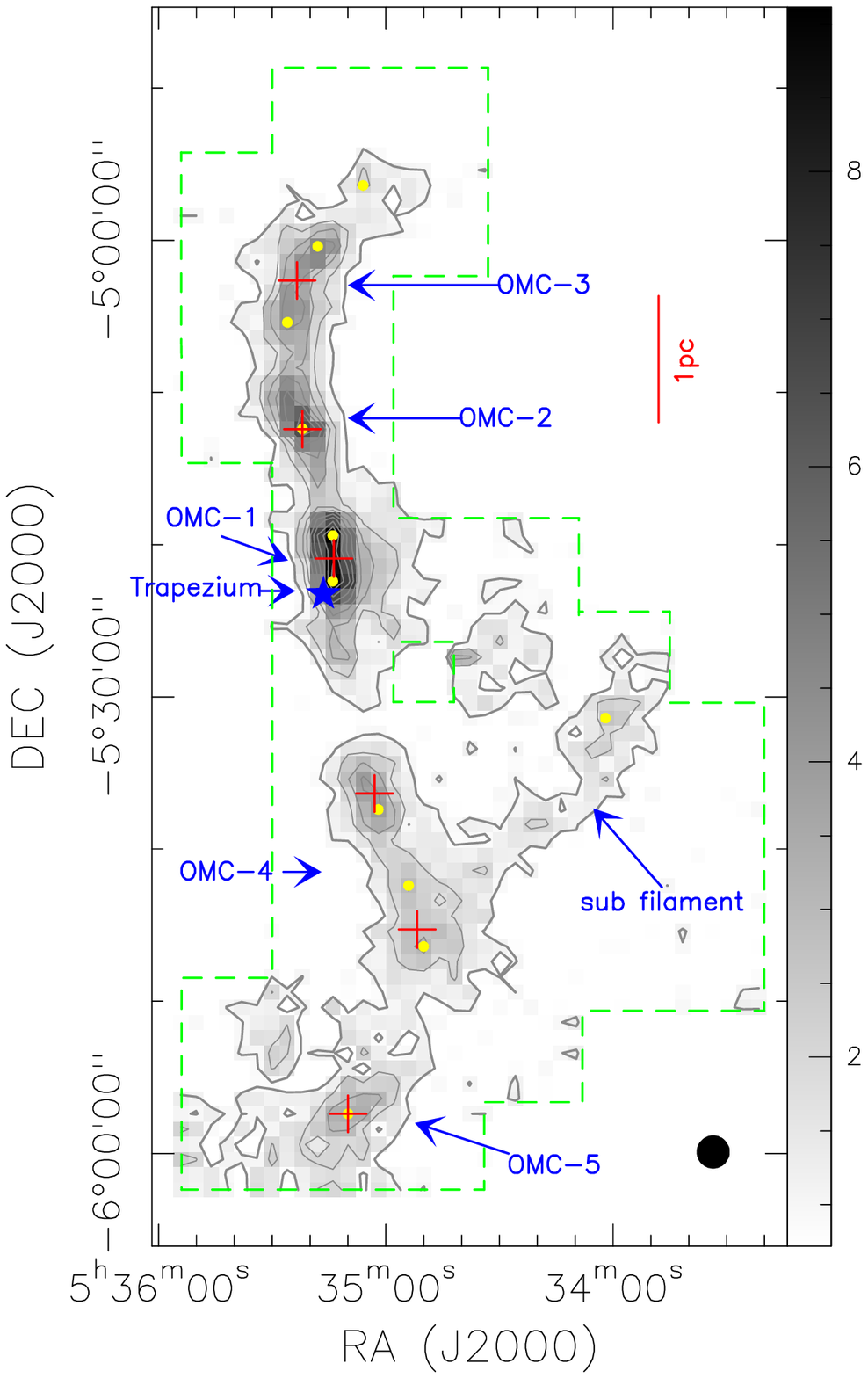}
   \caption{Left panel: False color image of the \object{ISF} (red for 500 $\mu$m, green for 250 $\mu$m, and blue for 100 $\mu$m, all Herschel data) overlaid with  NH$_{3}$ (1, 1) integrated intensity contours, as in Fig. \ref{FigMom0}. Right panel: "clumps"  and "sub-clumps" are labeled as red crosses and yellow filled circles, respectively, on the NH$_{3}$ (1, 1) integrated intensity map.}
   \label{FigHerschel}%
    \end{figure*}

   \begin{figure*}
   \centering
   \includegraphics[width=0.4\textwidth]{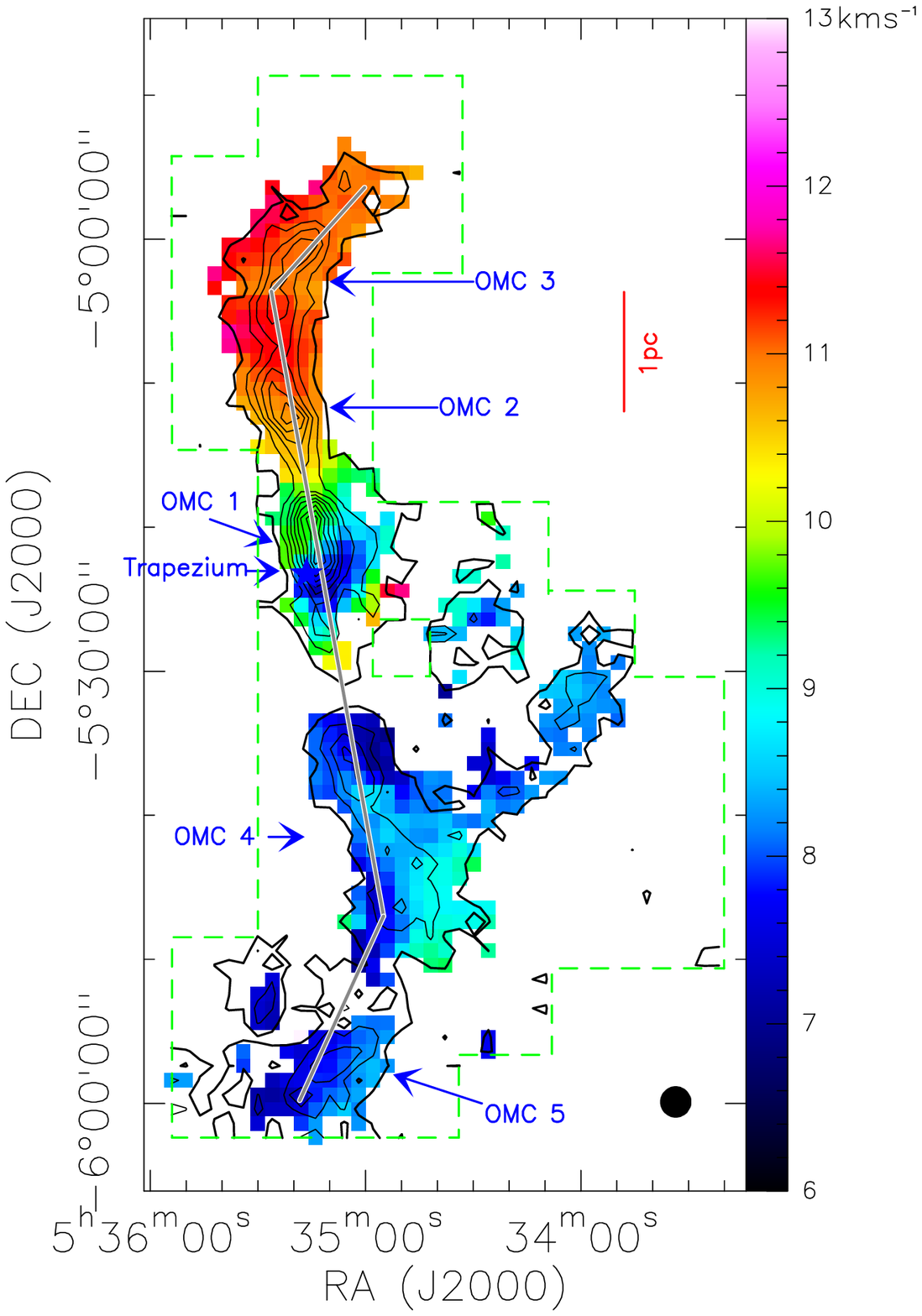}
   \includegraphics[width=0.4\textwidth]{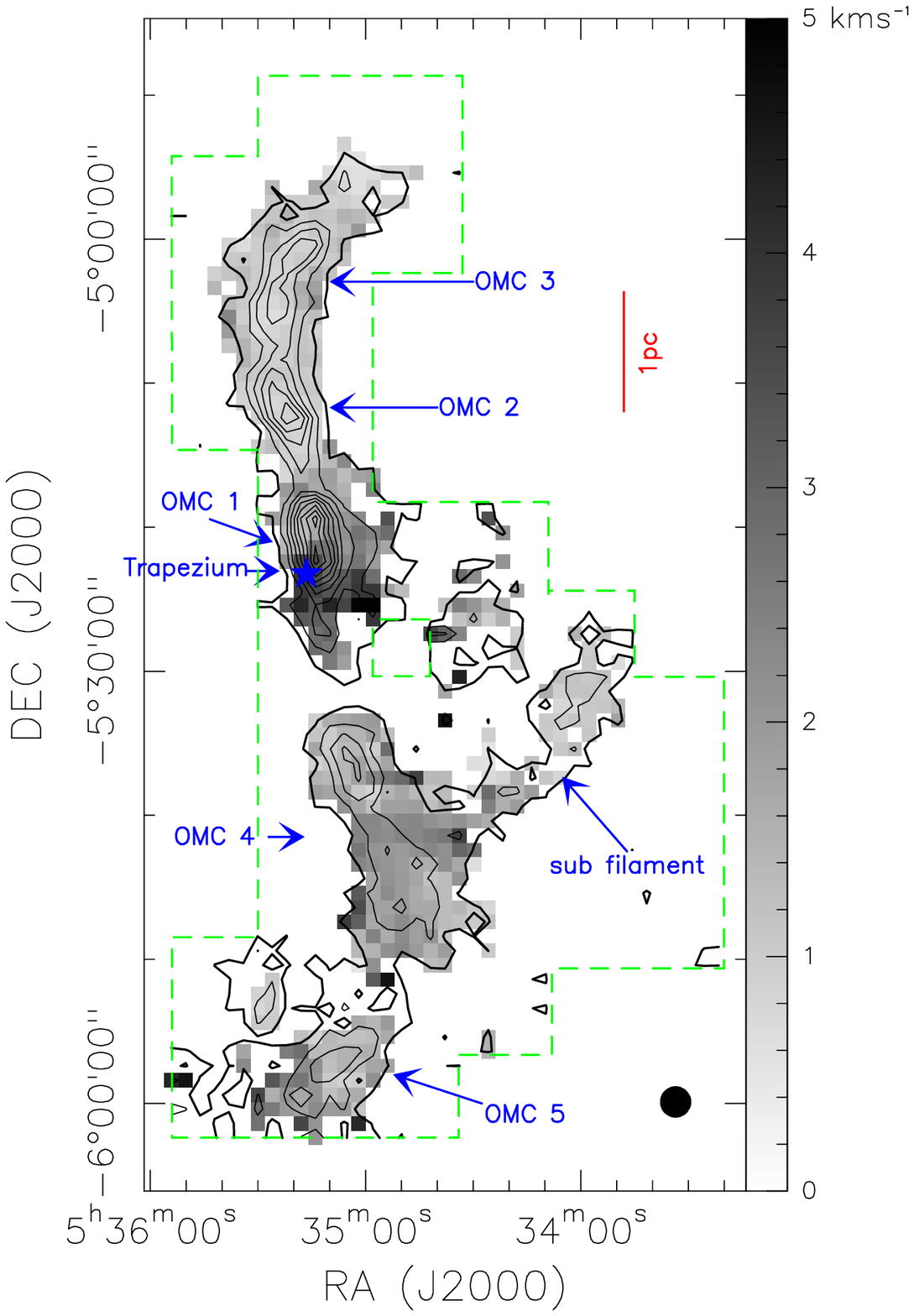}
   \caption{Kinematics of the \object{ISF} derived from NH$_{3}$ (1, 1). Left panel: Velocity map (color image) of  the \object{ISF}. Right panel: FWHM line width map (gray image) of  the \object{ISF}. Contours in each panel are the same as in the left panel of Fig. \ref{FigMom0}. The gray solid polyline in the left panel indicates the loci for the position-velocity diagram presented in Fig. \ref{FigPV}.}
   \label{FigVel}%
    \end{figure*}

   \begin{figure}
   \centering
   \includegraphics[width=\hsize]{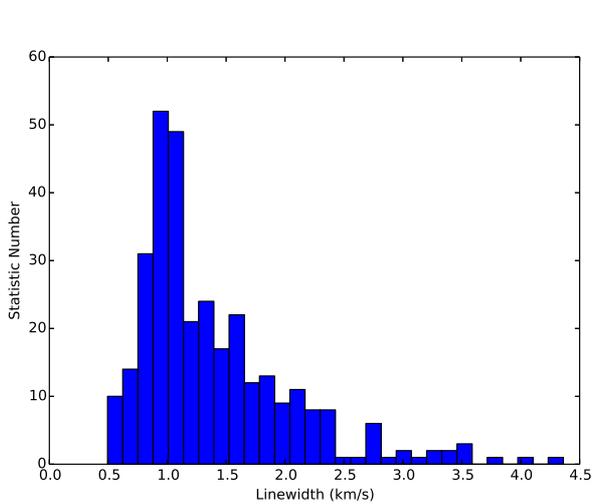}
    \caption{Histograms of the intrinsic FWHM line widths of the NH$_{3}$(1,1) main lines. These line widths referring to individual hyperfine components are derived from the GILDAS built-in `NH$_{3}$(1, 1)' fitting which can fit all 18 hyperfine components simultaneously.}
    \label{FigStaV}
   \end{figure}

   \begin{figure}
   \centering
   \includegraphics[width=\hsize]{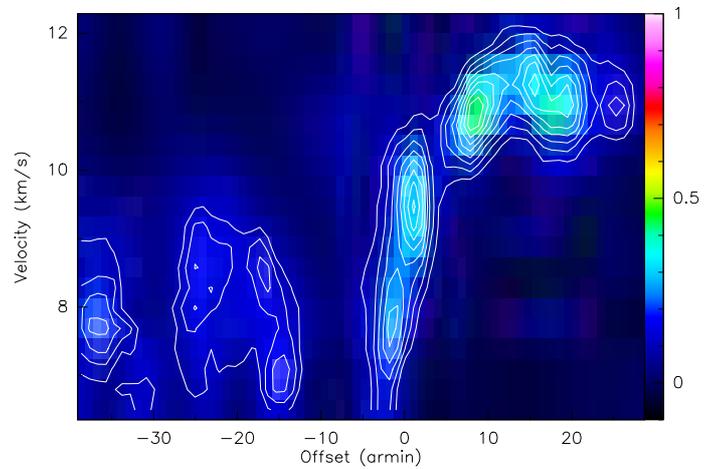}
    \caption{Position-velocity(PV) diagram of the NH$_{3}$ (1, 1) main line along the gray solid poly-line in the left panel of Fig. \ref{FigVel}. The cut runs through the \object{ISF} with negative offsets referring to the southern part of the filament. Offset 0 refers to the location of the Trapezium cluster. The threshold and step of the contours of the position-velocity diagram are all 10\% of the peak flux.}
    \label{FigPV}
   \end{figure}

   \begin{figure}
   \centering
   \includegraphics[width=\hsize]{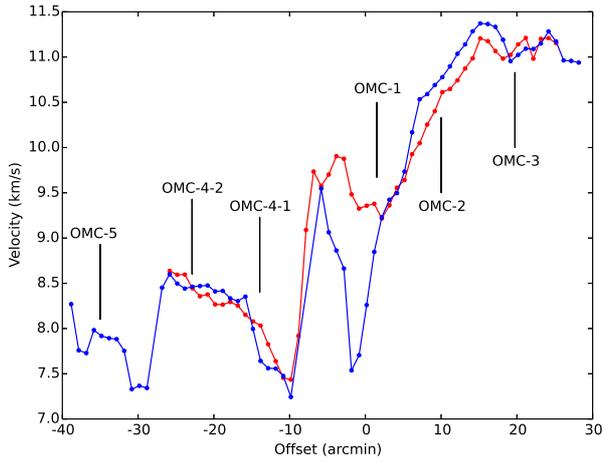}
    \caption{The centroid velocities of the NH$_{3}$(1,1) (blue dots and lines) and $^{13}$CO(1-0) (red dots and lines) averaged in 1'-width belts in declination along the \object{ISF}. Offset 0 refers to the location of the Trapezium cluster. The locations of OMC-1 to 5 are also displayed in this panel.}
    \label{FigVelDec}
   \end{figure}

   \begin{figure*}
   \centering
   \includegraphics[width=\hsize]{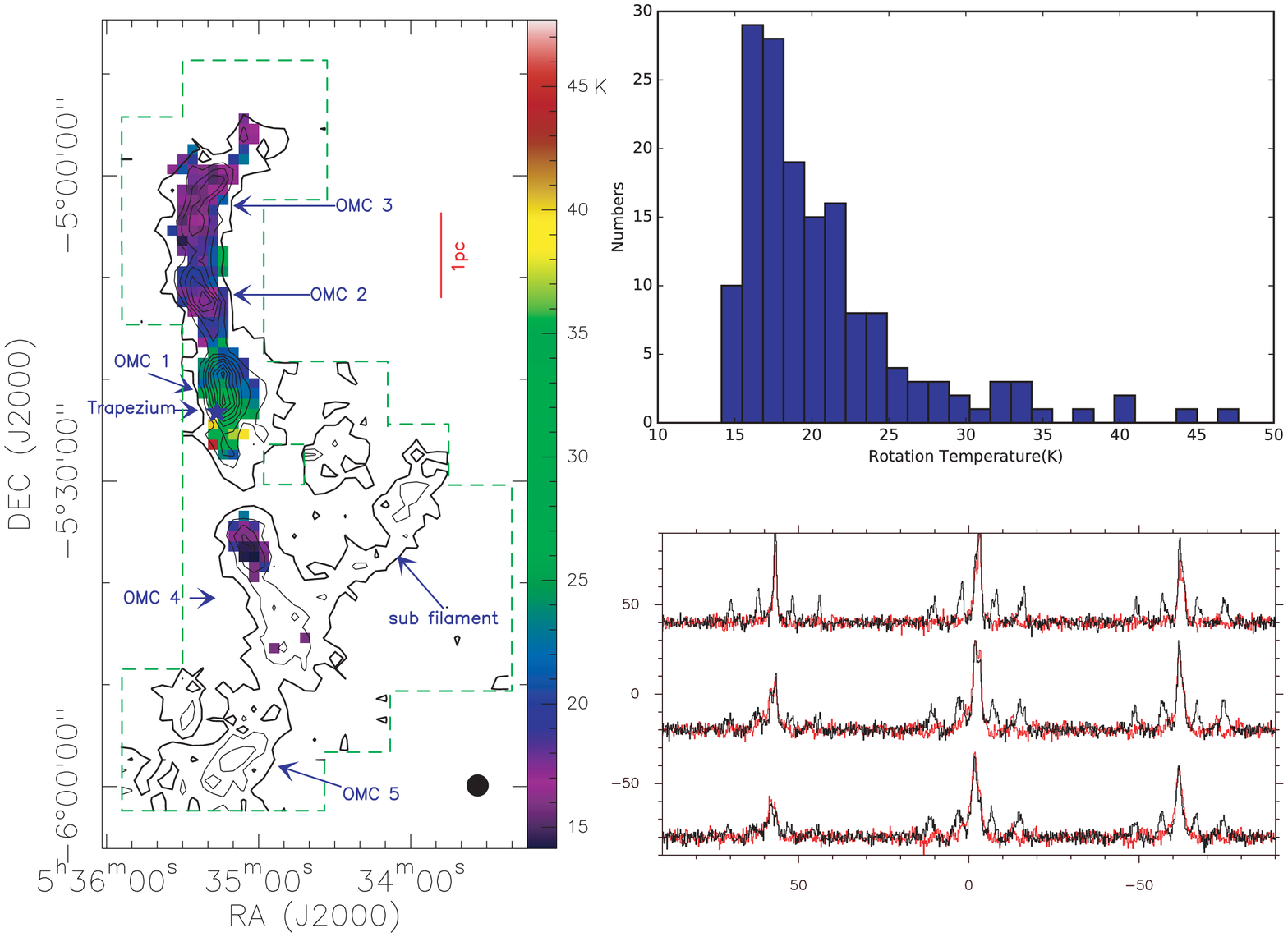}
   \caption{Left panel: Temperature distribution along the \object{ISF} overlaid with integrated intensity contours as in the left panel of Fig. \ref{FigMom0}. Most of the temperatures are about or below 20 K. Temperatures exceed 40 K at several positions. There the NH$_{3}$(2, 2) lines are even stronger than those of NH$_{3} $(1, 1) (see the lower right panel). Top right panel: Histogram of the rotation temperatures derived from NH$_{3}$. Lower right panel: The NH$_{3}$ (1, 1) (black profiles) and  NH$_{3}$ (2, 2) (red profiles) spectra around the Trapezium. The reference position corresponds to RA: 05:35:14, DEC: -05:22:24(J2000), the spacing is one arcminute.}
   \label{FigTem}%
   \end{figure*}

\begin{appendix} 
\section{Calibration stability}
\label{appCal}
\begin{figure}
\centering
\includegraphics[width=\hsize]{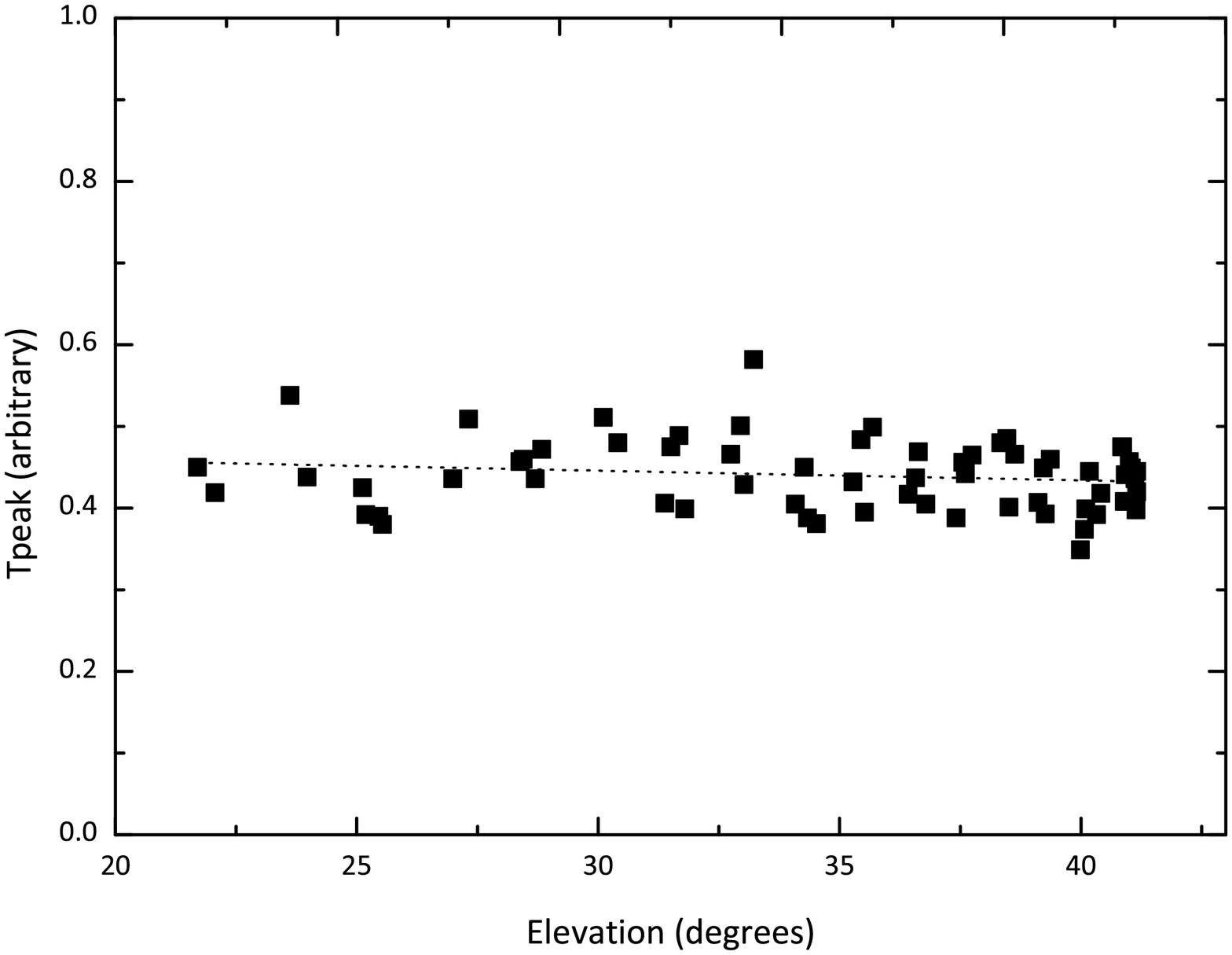}
  \caption{Uncorrected NH$_{3}$ (1, 1) main line intensities  against elevation of repeated observations toward the reference position. The position corresponds to RA: 05:35:14, DEC: -05:22:24 (J2000). The statistical error of the flux is about 9.5\%. The dotted line indicates a linear fit.}
  \label{FigRef}
\end{figure}

The system temperature was calibrated against a signal injected by a  noise diode, whose temperature is commonly determined by hot (ambient temperature) and cold (liquid nitrogen) loads. Because of a hardware problem, the output power of the diode noise source was abnormal but could maintain a constant value during the observations. To check the calibration stability, we observed the reference position (with (0, 0) offset)  every 2-3 hours. The spectral line fluxes turned out to be stable but no absolute flux calibration could be obtained from the data themselves. All observations of the reference position were made in the OTF mode of a small area of 3' $\times$ 3'. We regridded the data and then fitted the NH$_{3}$  (1, 1) main lines (the central group of NH$_{3}$ (1, 1) hyperfine components) to present the peak intensity distribution against elevation (Fig. \ref{FigRef}). We can see in Fig. \ref{FigRef} that there is no significant systematic variation and the dispersion is relatively small. The statistical error of the peak intensities is 9.5\%.

\begin{figure}
\centering
\includegraphics[width=\hsize]{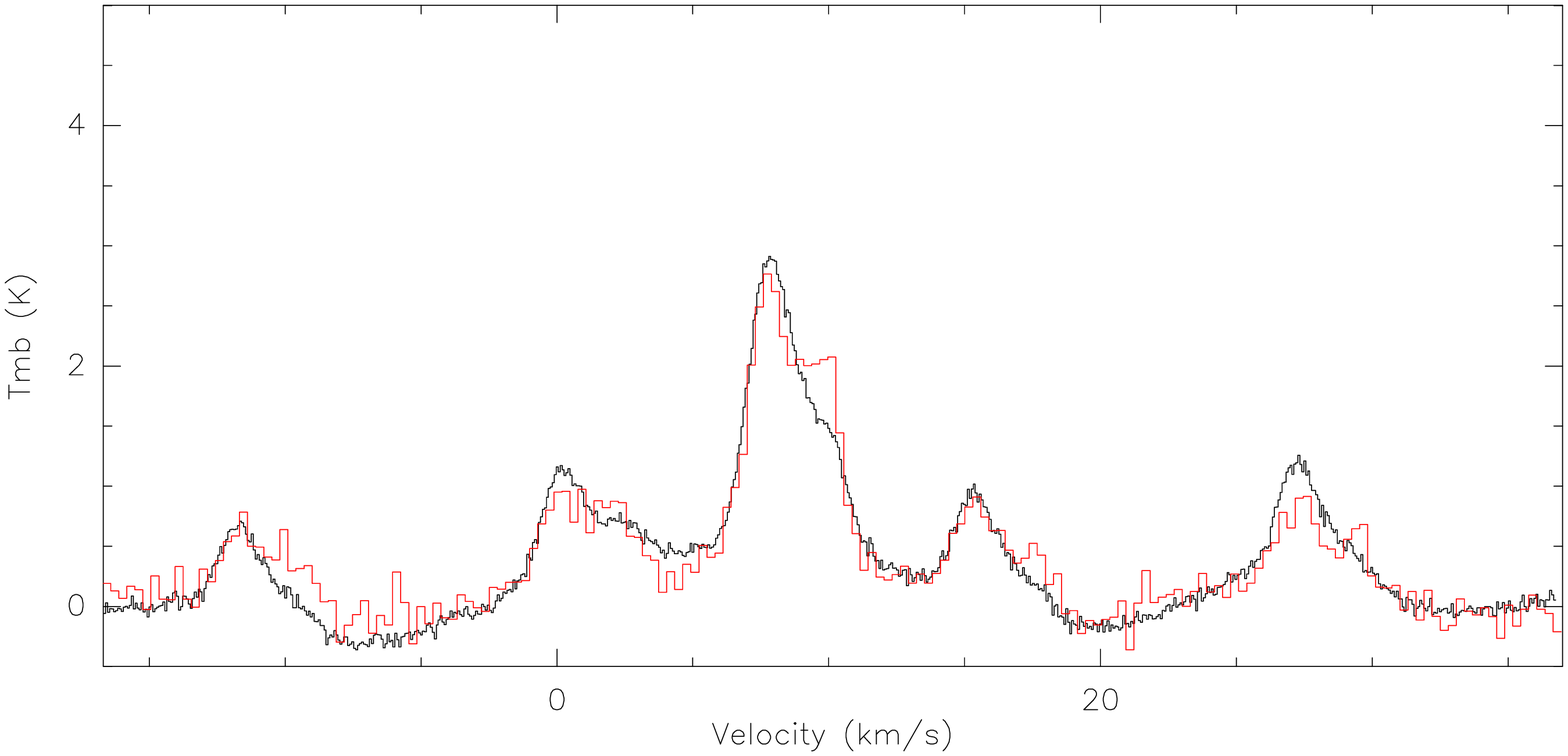}
  \caption{ The NH$_{3}$ (1, 1) spectra from the GBT (black) and our data (red, multiplied by 5.31) toward the reference position RA: 05:35:14, DEC: -05:22:24 (J2000).}
  \label{FigGBT}
\end{figure}

In order to calibrate our data to the main beam brightness temperature  ($T_{\rm mb}$) scale, we conducted a comparative analysis between the GBT  \citep[][]{2017ApJ...843...63F} and our NH$_{3}$ (1, 1) spectra. The GBT data were calibrated to $T_{\rm mb}$ scale with an estimated calibration uncertainty of about 10\% \citep[][]{2017ApJ...843...63F}. We smoothed the GBT data to our beamsize using the `XY\_MAP' routine in GILDAS. For the process our spectral cube was used as a template. We fitted the NH$_{3}$ (1, 1) main lines of the reference position and the derived peak line intensities are 2.46 K and 0.46 K for the GBT and our spectra respectively. Therefore our NH$_{3}$ data were all multiplied by the factor of 2.46 / 0.46 $\approx$ 5.31. The GBT spectrum (black) and our spectrum (red, multiplied by 5.31) of the reference position are illustrated in Fig. \ref{FigGBT}. We estimate our calibration uncertainty to be $\sqrt{(9.5\%)^{2} + (10\%)^{2}} \approx 14\% $.

\section{The sub-structures identified by the clumpfind algorithm}
\label{appSub}
\begin{figure}
\centering
\includegraphics[width=\hsize]{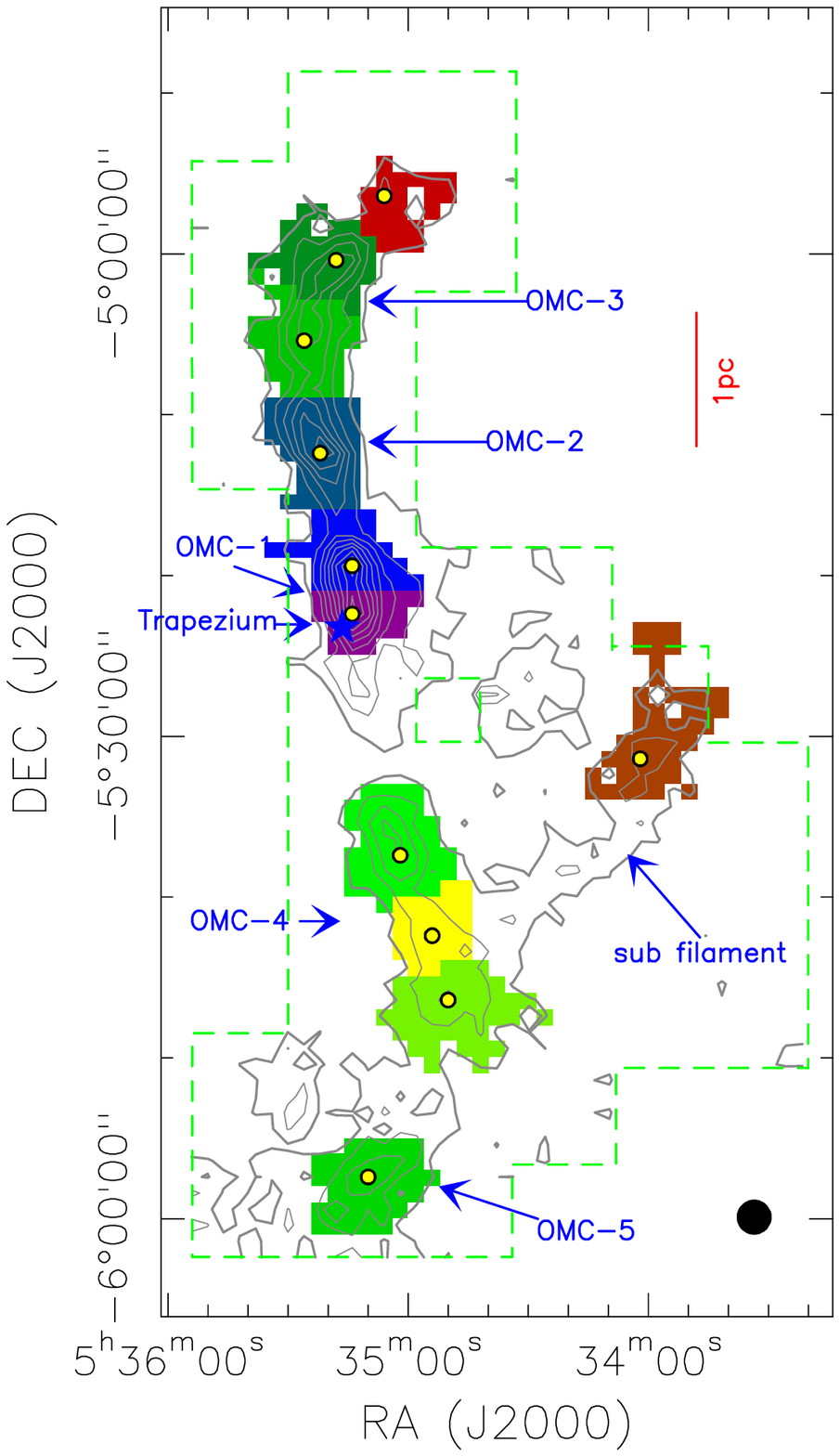}
  \caption{The colorized pixels illustrate the areas of the 11 sub-structures identified by the clumpfind algorithm. Contours are the same as in the left panel of Fig. \ref{FigMom0}. Yellow filled circles are same as in the right panel of Fig. \ref{FigHerschel}}
  \label{FigSub}
\end{figure}

To more clearly present the sub-structures identified by the clumpfind algorithm, we colorized the pixels used to define the sub-structures in Fig. \ref{FigSub}. We can see these sub-structures are all related to the emission peaks and are composed of a considerable number of pixels. We should note that there are three sub-structures identified in OMC-4.
The separation of the lower two sub-structures in OMC-4 may not be quite that obvious.
There is indeed an emission peak at the center of OMC-4 beside the northern and southern ones.

\section{Dissipation of turbulent energy}
\label{appDis}

\begin{figure*}
\centering
\includegraphics[width=0.8\textwidth]{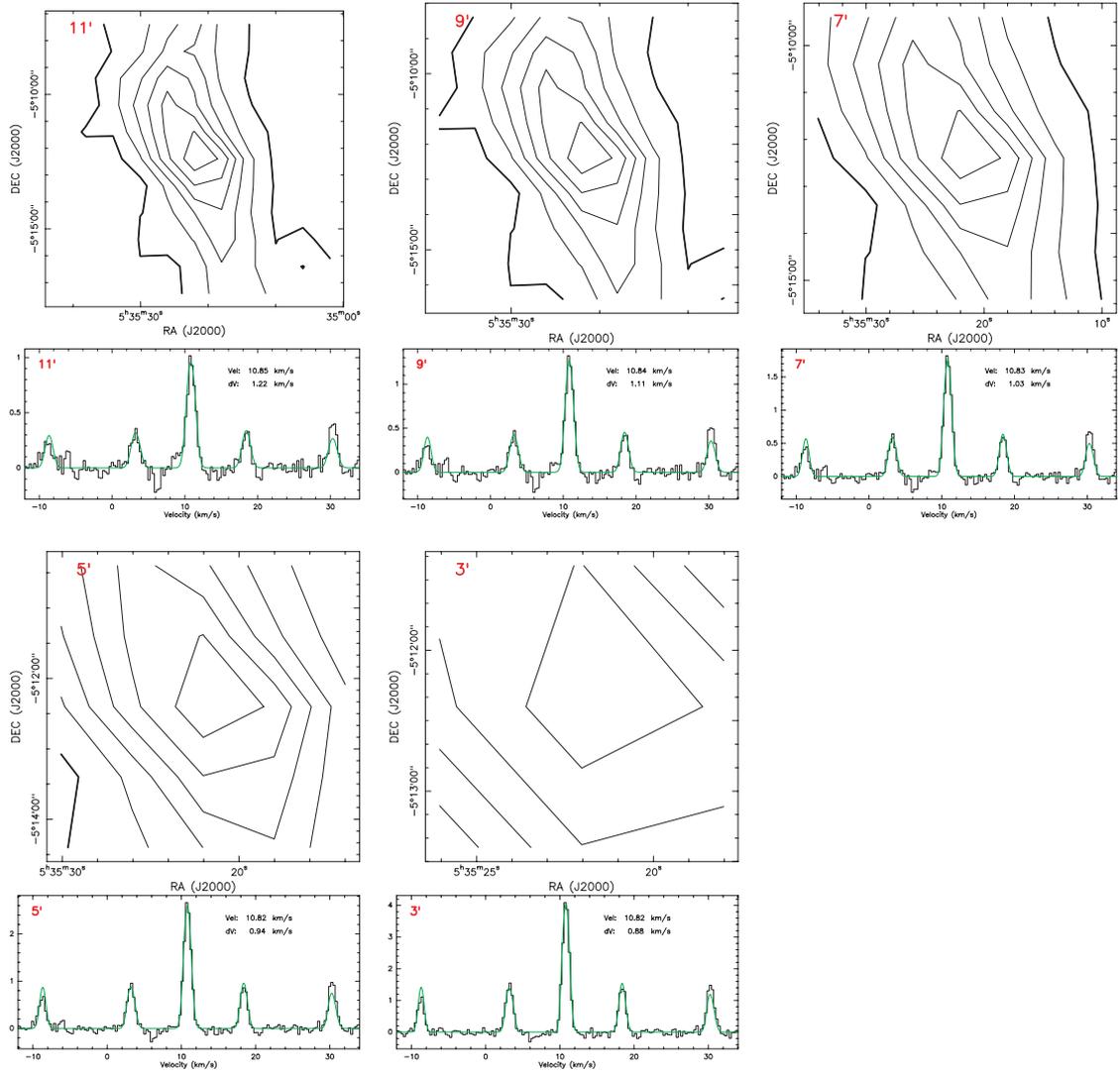}
\caption{From left to right and top to bottom, the contours are the integrated intensity (zeroth moment) maps of NH$_{3}$ (1, 1) with side lengths of 11', 9', 7', 5', and 3' respectively. These images are all centered at the red cross in OMC-2 illustrated in the right panel of Fig. \ref{FigHerschel}. The integration range of each panel is 6.5 < Vlsr < 12.5 km s$^{-1}$. Contours start at 0.96 K km s$^{-1}$ (4$\sigma$) and go up in steps of 0.96 K km s$^{-1}$. The spectra (black) overlaid with their fitting curves (green) are the averaged spectra within the regions corresponding to their upper panels. }
\label{FigOMC2}%
\end{figure*}

\begin{figure}
\centering
\includegraphics[width=\hsize]{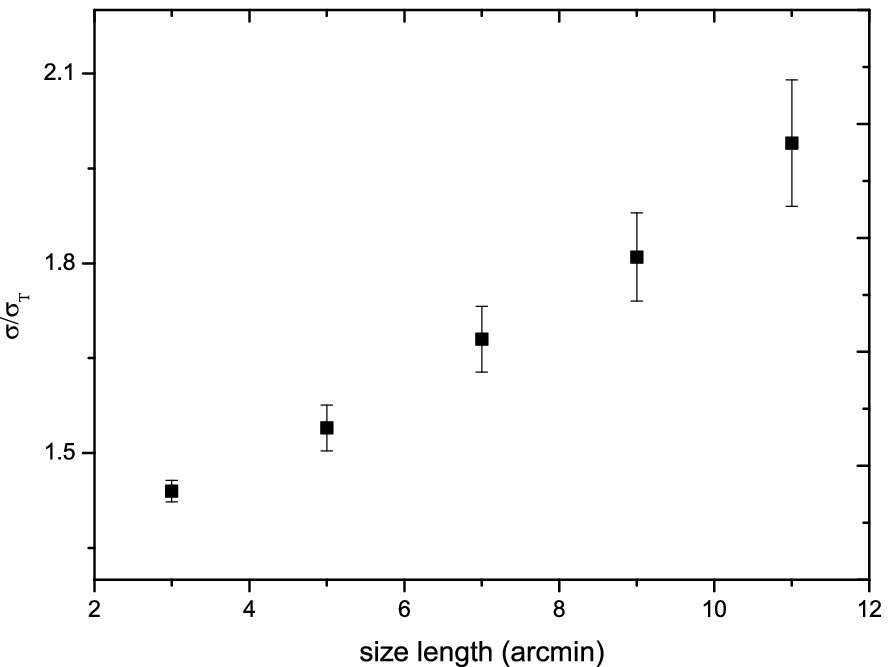}
\caption{The ratio of observed to thermal velocity dispersion, $\sigma_{V}/\sigma_{T}$, derived from the averaged spectra in Fig. \ref{FigOMC2} against the side length of the averaged regions.}
\label{FigOMC2dis}%
\end{figure}

Since the clump OMC-2 exhibits lower temperature and simpler structure (OMC-2 has only one sub-clump) than the other clumps, we used this relatively quiescent clump to roughly estimate the velocity dispersion at different scales. We averaged the spectra within squared regions with side lengths of 11', 9', 7', 5', and 3' respectively which are all centered at the red cross in OMC-2 illustrated in the right panel of Fig. \ref{FigHerschel} (see the images in Fig. \ref{FigOMC2}). We fitted the averaged spectrum using the GILDAS built-in `NH$_{3}$(1, 1)' fitting method (see the spectra in Fig.\ref{FigOMC2}). The derived intrinsic line widths referring to individual hyperfine components, $\Delta V_{obs}$, are 1.22$\pm$0.062 km s$^{-1}$, 1.11$\pm$0.046 km s$^{-1}$, 1.03$\pm$0.032 km s$^{-1}$, 0.94$\pm$0.022 km s$^{-1}$, and 0.88$\pm$0.017 km s$^{-1}$ respectively. All the results above are summarized in Table \ref{table:1}. The results are also displayed in Fig. \ref{FigOMC2dis} in the form of $\sigma_{V}/\sigma_{T}$ versus side length, where, $\sigma_{T}$ is the thermal velocity dispersion, which is commonly expressed in units of the the sound speed as c$_{s}$ = $\sqrt{kT/\mu_{H}m_{H}}$. The mean temperature of OMC-2 derived from the averaged spectra of NH$_{3}$(1, 1) and (2, 2) is about 20 K. We can see that as the size length decreases, the velocity dispersion decreases and this decrease is approximately linear.
Thus, the turbulence dissipates in the inner part of the \object{ISF} and  we speculate that  the thermal pressure of gas becomes more dominant.

We can see that in Fig. \ref{FigOMC2},  absorption features are present at 6 km s$^{-1}$ and 20km s$^{-1}$. In fact, the right side of the main line and all satellite lines present potential absorption features. However, the main line with broader line width covers the absorption feature. By checking the raw data, these features are not from  emission by the reference position. However, the S/Ns of these features are all lower than 1.8 and they are therefore not analyzed in this work.

\begin{table*}
\caption{Intrinsic (referring to individual hyperfine components) FWHM line widths, $\Delta V_{\rm obs}$, associated velocity dispersions, $\sigma_{\rm V}$, and ratios between these velocity dispersions and the thermal velocity dispersion for the NH$_{3}$ (1,1) data displayed in Fig. \ref{FigOMC2}.}            
\label{table:1}      
\centering                          
\begin{tabular}{c c  c c}        
\hline\hline                 
size(')  & $\Delta V_{obs}(km s^{-1})$  & $ \sigma_{V} (km s^{-1})$    & $\sigma_{V}/\sigma_{T}$\\    
\hline                        
   11& 1.22$\pm$6.15e-2   & 0.52$\pm$0.026   & 1.99$\pm$0.10\\      
   9 & 1.11$\pm$4.55e-2   & 0.47$\pm$0.019   & 1.81$\pm$0.07\\      
   7 & 1.03$\pm$3.20e-2   & 0.44$\pm$0.014   & 1.68$\pm$0.05\\
   5 & 0.94$\pm$2.22e-2   & 0.40$\pm$0.009   & 1.54$\pm$0.04\\
   3 & 0.88$\pm$1.69e-2   & 0.37$\pm$0.007   & 1.44$\pm$0.03\\

\hline                                   
\end{tabular}
\end{table*}

\section{The residuals of the velocity fitting }
\label{appRes}
\begin{figure}
\centering
\includegraphics[width=\hsize]{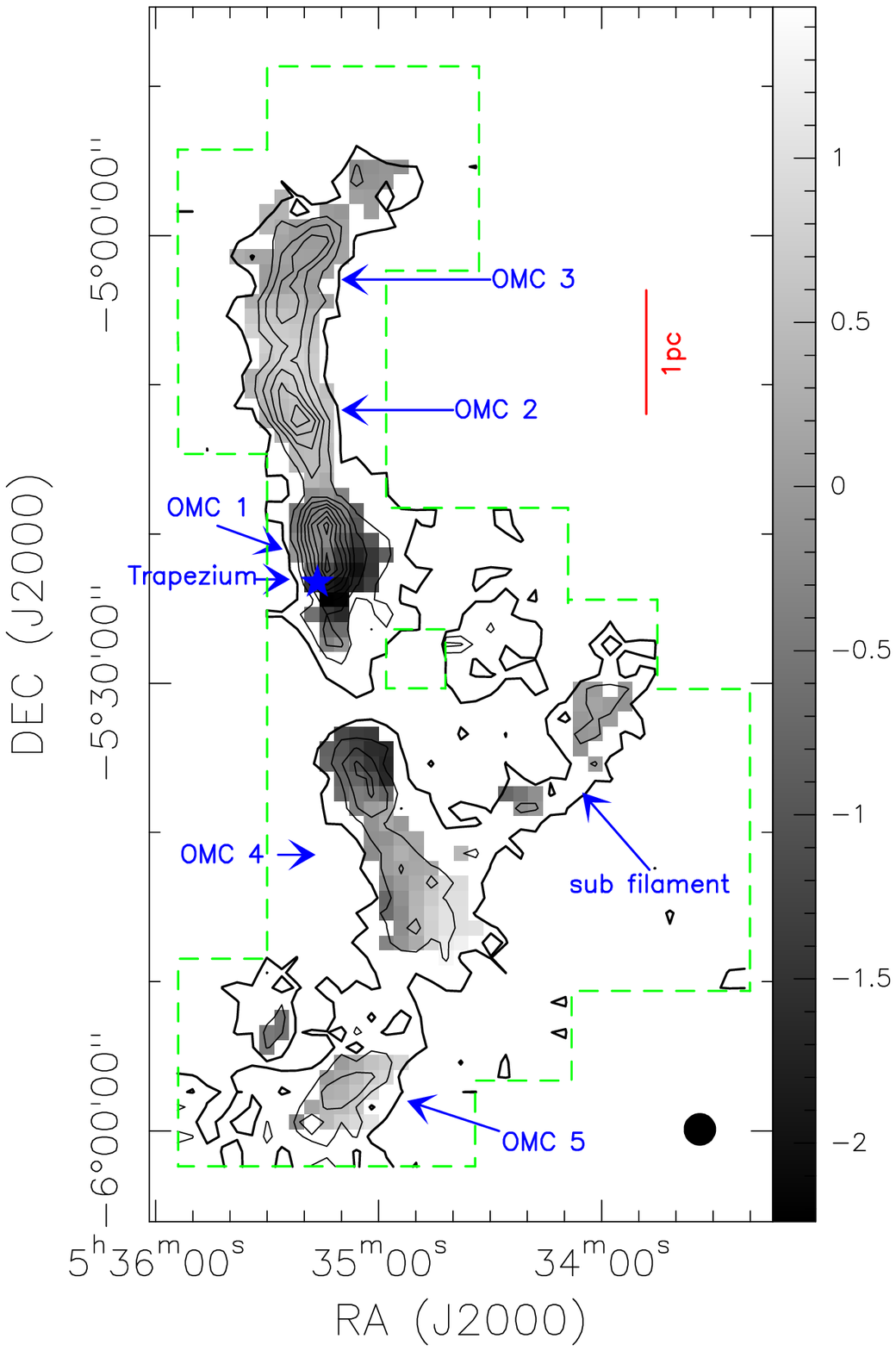}
  \caption{Velocity residual (V$_{\rm obs}$\,$-$\,V$_{\rm rigid}$) map (gray image) of the \object{ISF} derived from our NH$_{3}$. Contours are as in the left panel of Fig. \ref{FigMom0}.}
  \label{FigVelRe}
\end{figure}

\begin{figure}
\centering
\includegraphics[width=\hsize]{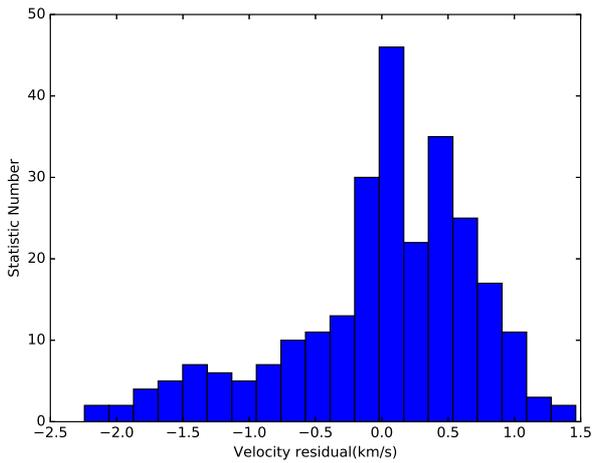}
\caption{Histograms of the velocity residuals (V$_{\rm obs}$\,$-$\,V$_{\rm rigid}$) derived from our NH$_{3}$.}
\label{FigVelReStat}
\end{figure}

The linear velocity fitting is based on the assumption that the \object{ISF} is a rigid body. To check this assumption, we present the distribution of the velocity residuals between the observed velocity and the fitting velocity in Fig. \ref{FigVelRe} and also the statistics of the velocity residuals in Fig. \ref{FigVelReStat}.
From Fig. \ref{FigVelRe}, we can see that most of the velocity residuals are distributed in the range V$_{\rm obs}$\,$-$\,V$_{\rm rigid}$ = $-1$\,km\,s$^{-1}$ to $1$\,km\,s$^{-1}$ which is even more clearly presented in Fig. \ref{FigVelReStat}. The larger velocity residuals are mainly negative  and mostly located at the southern part of OMC-1 and the northern part of OMC-4.
In general, the velocity residuals do not present major departures from a linear velocity distribution.

\end{appendix}

%
   \bibliographystyle{aa} 
   \bibliography{30316corr} 
%

\end{document}